# Burr, Lomax, Pareto, and Logistic Distributions from Ultrasound Speckle

**Kevin J. Parker,**[1*] **Sedigheh S. Poul**[2]
[1]Department of Electrical and Computer Engineering, University of Rochester
[2]Department of Mechanical Engineering, University of Rochester

***Corresponding Author:**
Kevin J. Parker**,** Department of Electrical and Computer Engineering, 724 Computer Studies Building, Box 270231, Rochester, NY, 14727-0231, USA.  E-mail: <u>kevin.parker@rochester.edu</u>

**Acknowledgments and funding:** This work was supported by National Institutes of Health grant R21EB025290.  The authors thank Terri Swanson and Theresa Tuthill of Pfizer Inc. for providing the RF data from their liver studies.  Thanks are also due to Professor Nicholas George for his profound insights on and contributions to speckle theory. The authors are grateful to Dr. R. James White and Gary Ge for the 3D vasculature data set and its rendering.

## Abstract

After 100 years of theoretical treatment of speckle patterns from coherent illumination, there remain some open questions about the nature of ultrasound speckle from soft vascularized tissues. A recent hypothesis is that the fractal branching vasculature is responsible for the dominant echo pattern from organs such as the liver. In that case an analysis of cylindrical scattering structures arranged across a power law distribution of sizes is warranted. Using a simple model of echo strength and basic transformation rules from probability, we derive the first order statistics of speckle considering the amplitude, the intensity, and the natural log of amplitude. The results are given by long tailed distributions that have been studied in the statistics literature for other fields. Examples are given from simulations and animal studies, and the theoretical fit to these preliminary data support the overall framework as a plausible model for characterizing ultrasound speckle statistics.

## Introduction

The study of speckle as a random interference phenomenon from coherent illumination is over 100 years old. The early work in light (predating the laser) utilized prisms to select a narrow band and study scattering[1], but even at that time the author said, "The theme of our investigation is an old one." With the advent of radar and laser sources, the research on the mathematical properties of optical speckle were accelerated.[2-8] In medical ultrasound, the mathematical treatment of speckle patterns is over 40 years in extent[9] and has developed into a rich set of models for the statistics of backscattered echoes from tissues[10].

For much of medical ultrasound, important goals include the differentiation of normal vs. pathological tissues, the detection of lesions, and the post-processing of B-scans for improved rendering of images including computer assisted diagnosis by algorithms. All of these tasks are strengthened by a careful analysis of speckle or texture from scatterers within normal soft tissues, and then any changes associated with pathological conditions. Accordingly, over time a number of models of ultrasound speckle have been postulated, and many of these models have been adapted from earlier work from optics and electromagnetics. These models include the classical Rayleigh distribution[9,11-13], the K-distribution[14-16], a Rician distribution[6,17,18], the Nakagami distribution[19-21], a "marked model" distribution[22,23], and other advanced models[24,25] with continuing applications to a variety of clinical targets[26-29].

Recently, we have proposed an alternative approach to the first and second order statistics of speckle from soft vascularized tissues[30-32]. Essentially, this model postulates that the fractal branching vasculature and fluid channels have an acoustic impedance mismatch of approximately 3% with respect to the surrounding tissue parenchyma. This mismatch forms the dominant set of inhomogeneities in normal soft vascularized tissues such as the prostate, thyroid, liver, and brain, and therefore the canonical scattering element is a cylinder, not a point or a sphere. Given the multi-scale, fractal structure of the vasculature, an ensemble average over all sizes from large to small leads to power law functions which propagate through different transfer functions and probability density functions (PDFs). This paper examines the first order statistics of speckle from tissue under the assumptions inherent in the framework where weak (Born approximation) scattering originates from a fractal branching set of cylindrical vessels within a reference medium and interrogated by a bandpass ultrasound pulse. It is shown that the echo amplitude, intensity, and log amplitude histograms can be modeled by conventional PDFs that are known in the statistics literature, and therefore have well described properties. These all contain a power law parameter that originates from the tissue structure itself. Preliminary examples from a 3D wave simulation of scattering and an animal imaging study of the liver are given to demonstrate the relevance of these functions.

## Theory

*Spatial Convolutions and Transforms*

The key assumptions and formulas used in deriving the first order statistics of speckle from a fractal branching vasculature are summarized below. First, we assume that a bandpass pressure pulse $P$ propagating in the $x$ direction with velocity $c$ can be approximated by separable functions[33]:

$$P\left(y, z, t - \frac{x}{c}\right) = G_y(y, \sigma_y) G_z(z, \sigma_z) P_x\left(t - \frac{x}{c}\right), \quad (1)$$

where $G_y(y, \sigma_y)$ is a Gaussian-shaped transverse beampattern in the $y$-direction with $\sigma_y$ representing the width parameter, similar for $G_z(y, \sigma_z)$, and $P_x$ is the propagating bandpass pulse shape in the $x$-direction.

Next, applying a 3D convolution model[33-35], we assess the *dominant* echoes from the pulse interacting with each generation of elements in a branching, fractal, self-similar set of vessels shown in **Fig. 1**, and whose number density as a function of radius $a$ follows a power law behavior[36] $N(a) = N_0/a^b$, where $b$ is a real number greater than 1 defining the branching behavior of the fractal vascular tree and $N_0$ is a constant determining the overall number density.

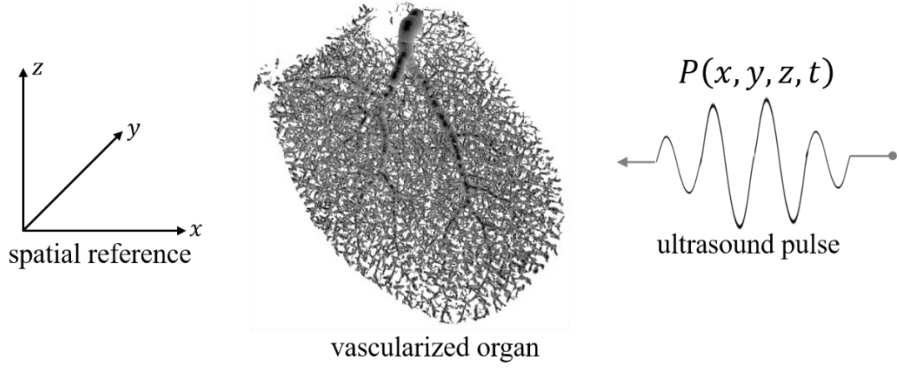

**Fig. 1** Model of 3D convolution of a pulse with the fractal branching cylindrical fluid-filled channels in a soft tissue.

The canonical scatter shape for any branch is a long fluid-filled cylinder of radius $a$ with long axis aligned along the $z$ - direction:

$$f(r) = \begin{cases} \kappa_0 & r \leq a \\ 0 & r > a \end{cases}$$

$$F(\rho) = \frac{\kappa_0 \cdot a \cdot J_1[2\pi a \cdot \rho]}{\rho}, \qquad (2)$$

where $\kappa_0$ is the fractional variation in compressibility, assumed to be $\ll 1$ consistent with the Born formulation, $F(\rho)$ represents the Hankel transform using Bracewell's convention[37], which is the 2D Fourier transform of the radially symmetric function $f(r)$, $J_1[\cdot]$ is a Bessel function of the first kind of order 1, and $\rho$ is the spatial frequency equal to $\sqrt{k_x^2 + k_y^2}$. The fractional variation in compressibility, $\kappa_0$, between blood vessels and liver parenchyma has been estimated to be approximately 0.03, or a 3% difference based on published data[31].

In addition, we also consider a "soft-walled" cylindrical vessel representing a less sharp transition in acoustic impedance between the fluid interior and the outer "solid" tissue:

$$f(r) = \frac{\kappa_0 \mathrm{e}^{-2\pi\sqrt{\left(\frac{r}{a}\right)^2 + 1}}}{\sqrt{\left(\frac{r}{a}\right)^2 + 1}} \text{ for } r > 0 \text{ and } a > 0. \tag{3}$$

Its Hankel transform is given by theorem 8.2.24 of Erdélyi and Bateman[38]:

$$F(\rho) = \frac{\kappa_0 a^2 \mathrm{e}^{-2\pi\sqrt{(a\rho)^2 + 1}}}{2\pi\sqrt{(a\rho)^2 + 1}} \text{ for } \rho > 0 \text{ and } a > 0. \tag{4}$$

The convolution of the pulse with a cylinder of radius $a$ is dominated by the case where the cylinder is perpendicular to the direction of the forward propagating pulse, the $x$-axis in our case. Thus, assuming an optimal alignment, the 3D convolution result is given by the product of the transforms:

$$^{3D}\mathfrak{I}\{\mathrm{echo}(x,y,z)\} = \mathfrak{I}^{3D}\{p(x,y,z)\} \cdot (k_x)^2 \, \mathfrak{I}^{3D}\{\mathrm{cylinder}(x,y,z)\}, \tag{5}$$

where the $(k_x)^2$ term pre-multiplying the cylinder transform stems from the Laplacian spatial derivative in the Born scattering formulation[39,40] and in the 3D convolution model[35,41].

By Parseval's theorem, the integral of the square of the transform equals the integral of the square of the echo, and provides a measure of the energy within the echo:

$$\iiint \{\mathrm{echo}(x,y,z)\}^2 \, dxdydz =$$
$$\kappa_0^2 \int_{kx=-\infty}^{\infty} \int_{ky=-\infty}^{\infty} \int_{kz=-\infty}^{\infty} \left(\mathfrak{I}^{3D}\{p(x,y,z)\}\right)^2 \cdot (k_x^2)^2 \cdot \left(\mathfrak{I}^{3D}\{\mathrm{cylinder}(x,y,z)\}\right)^2 dk_x dk_y dk_z. \tag{6}$$

We assume the left side of eqn (6) is also proportional to the average intensity $I$ of the echo as a function of the deterministic parameters on the right side, and the square root of this is proportional to the amplitude of the echo. From numerical evaluations of eqn (6) using either of two cylinders (eqn (2) or (3)) and either of two bandpass pulses (Gaussian Hermite or hyperbolic secant) we found[32] an approximation which will be useful for deriving a closed form solution of the echo amplitude $A$, $A[a] = A_0\sqrt{a - a_{min}}$ for $a > a_{min}$, and 0 otherwise. This approximate relationship is justified by the nearly linear increase in the energy term above some minimum threshold, and the asymptotic modulus of $J_1(ak)$ which is proportional[42] to $\sqrt{2/(\pi ak)}$ as $ak$ becomes large. The exact shape is dependent on the particular pulse shape's spectrum and the beampattern.

So as a working approximation, we apply the relation $A[a] = A_0\sqrt{a - a_{min}}$ (or for intensity $I$, $I[a] = I_0(a - a_{min})$) for $a > a_{min}$. The parameter $a_{min}$ depends on a number of factors, including the dynamic range selected (for example, 45 dB) and the Rayleigh scattering (long wavelength, small $a$) behavior of the cylinder interacting with the particular pulse transmit signal, along with the noise floor and quantization floor of the receiver.

## *Probability of Amplitudes*

Consistent with fractal models[36,43], we assume that along the line of propagation of the incident pulse in Figure 1, and within the interrogated ensemble, the probability density of encountering vessels at different radii given by a power law:

$$p(a) = ((b-1)/a_{min})(a/a_{min})^{-b} \tag{7}$$

for $a > a_{min}$ and $b > 1$, and this will be transformed[44] into the probability distribution of amplitudes, $p(A)$. The general transformation rule is:

$$p[A] = \frac{1}{dA/da} p[a]. \tag{8}$$

In our case, the derivative $dA/da = \left[(1/2)A_0\right]/\sqrt{a - a_{min}}$, and the inverse function is $a[A] = (A/A_0)^2 + a_{min}$. Thus, substituting these into eqn (8) the PDF $p[A]$ is:

$$p[A] = \frac{2A(a_{min})^{b-1}(b-1)}{A_0^2 \left[\left(\frac{A}{A_0}\right)^2 + a_{min}\right]^b}. \tag{9}$$

Furthermore, by substituting $\lambda = A_0\sqrt{a_{min}}$, we find this reduces to a two-parameter distribution:

$$p[A] = \frac{2A(b-1)}{\lambda^2 \left[\left(\frac{A}{\lambda}\right)^2 + 1\right]^b}, \tag{10}$$

which is a <u>Burr Type XII</u> distribution[45,46] with $c = 2$. Thus, the speckle PDF is given by a two-parameter distribution with known analytic expressions for its cumulative distribution function, and moments[46]. For example, the peak of the distribution occurs at $A = \lambda/\sqrt{2b-1}$ for $b > 1/2$.

Thus, the Burr distribution (10) describes the expected histogram distribution of echo amplitudes from a fractal branching set of Born cylinders. In particular, the power law parameter $b$ is a major parameter of interest.

*Probability of Intensity*

For completeness, we examine the PDF of echo intensity from this model. Again, assume that the probability distribution of a fractal branching vasculature is described by a power law in radius $a$ ($a_{min}$ now pertains to vessel size minimum) as given in eqn (7).

Furthermore, assume the average backscatter intensity $I(a) \cong I_0(a - a_{min})$ for $a > a_{min}$, zero otherwise. Then,

$$p_I(I) = \frac{p(a)}{\left(\frac{dI}{da}\right)}, \tag{11}$$

where

$$\frac{dI}{da} = I_0; \quad a = \left(\frac{I}{I_0}\right) + a_{\min}. \tag{12}$$

Thus, using the transformation rules:

$$p_I(I) = \frac{\left(\frac{b-1}{a_{\min}}\right)\left(\frac{a}{a_{\min}}\right)^{-b}}{I_0} = \frac{(b-1)}{(a_{\min})^{1-b} I_0 a^b} = \frac{(b-1)}{(a_{\min})^{1-b} I_0 \left(\frac{I}{I_0} + a_{\min}\right)^b}, \tag{13}$$

which is a <u>Lomax</u> distribution, also related to a <u>Pareto type II</u> distribution[47]. This can be more compactly written as:

$$p_I(I) = \frac{(b-1)\lambda_2^{b-1}}{(I+\lambda_2)^b} \tag{14}$$

for $I > 0$ and $b > 1$, and where $\lambda_2 = I_0 \cdot a_{\min}$.

## *Probability of Log-Transformed Envelope*

In ultrasound imaging, it is conventional to display the echo amplitudes using a log or dB scale to help with visualization of the wide dynamic range. The log transformation affects the distribution, and again using probability transformation rules[44], let $y = \ln(A)$, $dy/dA = 1/A$, and $A = \mathbf{e}^y$. Then:

$$p_y(y) = \frac{p(A)}{dy/dA} = \frac{2(b-1)\exp[2y]}{\lambda^2 \left(\dfrac{\exp[2y]}{\lambda^2} + 1\right)^b} \tag{15}$$

for $y > 0$ and $b > 1$. Note that in probability literature, this is related to the <u>generalized logistic distribution</u>, and for the special case where $\lambda = 1$ and $b = 2$ this becomes the sech$^2$ distribution[48].

Thus, each form of the received echoes from speckle {amplitude, intensity, ln(amplitude)} are given by standard PDFs known in the literature {Burr type XII, Lomax, generalized logistic}. These have now been derived based on a simple transformation of probability distributions using a mapping function linking vessel radius to echo strength. However, when multiple vessels are present within the interrogating pulse, then more complex treatment is required.

*Increasing Power Law b with Complex Summation*

Let us assume that the interrogated sample volume in an imaging system is large enough to encompass two or several discrete cylindrical scatterers simultaneously. Because of the RF modulation of the pulse, their echo amplitudes will be complex. Because of the fractal distribution, the probability distribution of each individual reflected echo amplitude has already been given as a Burr distribution. Note that historically, the Rayleigh distribution was derived by considering a complex summation of many independent point scatterers, then by invoking the central limit theorem a Gaussian distribution is generated from the sum of many identical and independent random variables[48-52]. In marked contrast in our case, we have cylindrical scatterers from a power law probability distribution over a wide range of radii, and we do not anticipate having so many vessels within a sample volume that we can invoke the central limit theorem. Furthermore, power law distributions (long tail distributions) have slow convergence to the central limit, and so it is instructive to look at the complex sum of two or few scatterers. The statistics literature has derived the sum of random variables of these distributions but the solutions generally involve complicated generalized functions or series[47,48,53-57]. To simplify this, we examine a complex Burr summation.

With reference to **Fig. 2**, using standard notation we can write the amplitude of the complex sum of two phasors $A$ and $B$ as:

$$|C| = \sqrt{\left(|A|\cos\theta_A + |B|\cos\theta_B\right)^2 + \left(|A|\sin\theta_A + |B|\sin\theta_B\right)^2}, \tag{16}$$

where $|A|$ and $|B|$ are independent Burr-distributed amplitudes (sampled from the echoes returning from the fractal branching network), and $\theta_{A,B}$ are independent and uniformly distributed over $0 \leq \theta < 2\pi$.

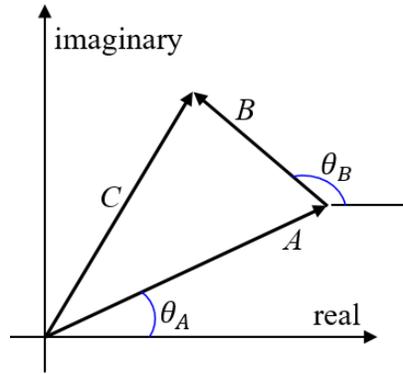

**Fig. 2** The vector sum of two independent phasors, pertaining to the real and imaginary parts of a complex addition as is commonly found in models of scattering.

The new random variable $|A|\cos\theta_A$ is given by the product distribution law[44,54] involving an integral over the PDFs for both $|A|$ and $\cos\theta_A$. By the transformation rule we can easily show that if $p(\theta) = 1/2\pi$ where $0 \leq \theta \leq 2\pi$, then $p(y = \cos\theta) = 1/\left(2\pi\sqrt{1-y^2}\right)$ where $-1 < y < 1$.

Then, if $\hat{A} = |A| \cdot y$, the product distribution yields:

$$p(\hat{A}) = \int_{-\infty}^{\infty} p_A(A) p_y\left(\frac{z}{A}\right)\frac{1}{|A|}da$$
$$= \int_{A=Z}^{\infty} \frac{4(b-1)}{2\pi(A^2+1)^b \sqrt{1-\left(\frac{z}{a}\right)^2}} dA \qquad (17)$$
$$= \frac{(b-1)\Gamma\left(b-\frac{1}{2}\right)}{\sqrt{\pi}\Gamma[b](1+\hat{A}^2)^{b-\frac{1}{2}}}$$

for $b > 1$ and $\lambda = 1$, where $\Gamma$ is the gamma function and thus $p(\hat{A})$ is a double-sided Pearson PDF with $E[\hat{A}] = 0$.

Next we need the PDF for $|A|\cos\theta_A + |B|\cos\theta_B$. The sum of independent and identically distributed (IID) variables is given by the convolution formula[44]. We found closed form solutions as ratios of polynomials for convolutions of eqn (17) only for integer orders of $b - 1/2$; it is instructive to look at one practical example. Let $Z_C = Z_A + Z_C$ with a PDF of $Z_{A,B}$ given by eqn (17), and let $b = 2.5$. Then we find:

$$p(Z_C) = \frac{4\pi(20 + Z_C^2)}{(4+Z_C^2)^3} = 4\pi\left(\frac{20}{(4+Z_C^2)^3} + \frac{Z_C^2}{(4+Z_C^2)^3}\right). \qquad (18)$$

The latter form emphasizes the important term with numerator 20, which dominates for $Z_C^2 < 20$ and in this example has denominator power of 3 (or more generally $b + 1/2$) Thus, the PDF of the sum of two Burr phasors' real parts is dominated by an increased power law of $b + 1/2$. In other words, the dominant power law term in the PDFs increases from 2.5 to 3. As more phasors are added, by induction this leading term increases. Specifically, a third phasor $Z_C$ leads to another convolution of the IID PDFs, raising the denominator power to 4 (or $b+3/2$). The final PDF of $|C|$ in eqn (16) requires further calculations and becomes complicated, however the trend towards increasing $b$ with increasing number of cylinders is revealed by the examination of the real part of the phasor addition as given above. These PDFs are summarized in **Table 1**.

**Table 1** Summary of soft tissue speckle PDFs.

| Measured signal | PDF | Formula |
|---|---|---|
| echo amplitude $A$ | Burr (XII) | $p(A) = \dfrac{2A(b-1)}{\lambda^2 \left[ \left(\dfrac{A}{\lambda}\right)^2 + 1 \right]^b}$ |
| echo intensity $I$ | Lomax (Pareto II) | $p_I(I) = \dfrac{(b-1)\lambda_2^{b-1}}{(I+\lambda_2)^b}$ |
| log intensity $y = \ln(A)$ | Generalized logistic (II) | $p_y(y) = \dfrac{2(b-1)\exp[2y]}{\lambda^2 \left( \dfrac{\exp[2y]}{\lambda^2} + 1 \right)^b}$ |
| phasor $\hat{A} = |A|\cos\theta_A$ | Pearson | $p(\hat{A}) = \dfrac{(b-1)\Gamma\left(b-\dfrac{1}{2}\right)}{\sqrt{\pi}\Gamma[b]\left(1+\hat{A}^2\right)^{b-\frac{1}{2}}}$ |
| sum of independent components $Z_C = A\cos\theta_A + B\cos\theta_B$ | mixed | $p(Z_C) = \dfrac{4\pi\left(20+Z_C^2\right)}{\left(4+Z_C^2\right)^3}$ |
| fractal distribution of branches radius $a$ | power law (Pareto) | $p(a) = \left((b-1)/a_{\min}\right)\left(a/a_{\min}\right)^{-b}$ |

## Methods

*Simulations*

In this study, to make a simple model of the liver parenchyma having vessels with fractal branching nature, a 3D block including multiscale cylindrical branches was generated to simulate the wave propagation and obtain the statistics of speckles. The block dimensions are 15 mm × 13 mm × 3 mm in the axial $(x)$, lateral $(y)$, and transverse $(z)$ directions, respectively, with the uniform grid element size of 69.4 $\mu$m approximately in all directions. The distribution of the cylindrical branches as scatterers with different radii obeys the power law behavior of eqn (7) with $b=2.5$. The cylindrical scatterers' radii ranged from 1 to 6 grid elements, and are randomly distributed in the background with no overlap among any two generated branches.

The k-Wave toolbox in MATLAB (The Mathworks, Inc., Natick, MA, USA) is employed to simulate the propagation of compressional waves in the time domain. This open-source toolbox uses the k-space pseudospectral approach to solve the acoustic wave equations.[58]

Using a virtual linear array transducer defined as the source and sensor in the k-Wave toolbox, an excitation signal is applied in the form of two transient toneburst cycles with a frequency of 4 MHz. This frequency is selected to lie in the common frequency range used for adult human abdominal scanning. For the material properties assignment, the speed of sound is set to 1540 m/s and 1500 m/s for the background and scatterers, respectively, and a uniform density of 1000 kg/m$^2$ is assumed for the entire medium with a small absorption coefficient. Moreover, in order to avoid the reflection effect from the boundary, the 3D domain is surrounded by an absorbing boundary layer, known in the k-Wave toolbox as a perfectly matched layer, which absorbs acoustic waves at the boundaries and minimizes reflection back to the domain.

A larger study focused on the effect of the number of scatterers per unit volume from these simulations was recently completed[59] across a range of parameters including $2<b<3$ and further details of the simulation can be found therein.

*Liver Scans*

Separately, experimental results were obtained from liver experiments. Rat experiments were reviewed and approved by the Institutional Animal Care and Use Committee of Pfizer, Inc. Groton Connecticut, where the ultrasound RF data were acquired using a Vevo 2100 (VisualSonics, Toronto, CA) scanner and a 21-MHz center frequency transducer (data provided courtesy of Terry Swanson). For the purpose of examining speckle PDFs, two scans (one normal and one fibrotic) were selected for having good quality B-scan images with adequate liver ROIs. The focal depth was set to 11 mm and positioned to the lower half of the liver in the sagittal plane. In analyzing the results from simulations and liver scans, parameter estimation was performed using MATLAB nonlinear least squares minimization of error, for two-parameter fits of the Burr distribution to the data.

## Results

An example from the simulation results are given in **Fig. 3**. **Fig. 3(a)** shows a 3D orientation of the transducer and random cylindrical scatterers in the domain. Only a few branches are shown here to clarify their orientation as perpendicular to the axial propagation of the interrogating pulse. **Fig. 3(b)** illustrates one realization of a random distribution of weak cylindrical scatterers of various diameters following a power law (fractal) function with $b = 2.5$ and $N_0 = 250$. **Fig. 3(c)** shows the resulting 4 MHz B-scan formed from the reflected echoes showing a characteristic speckle pattern.

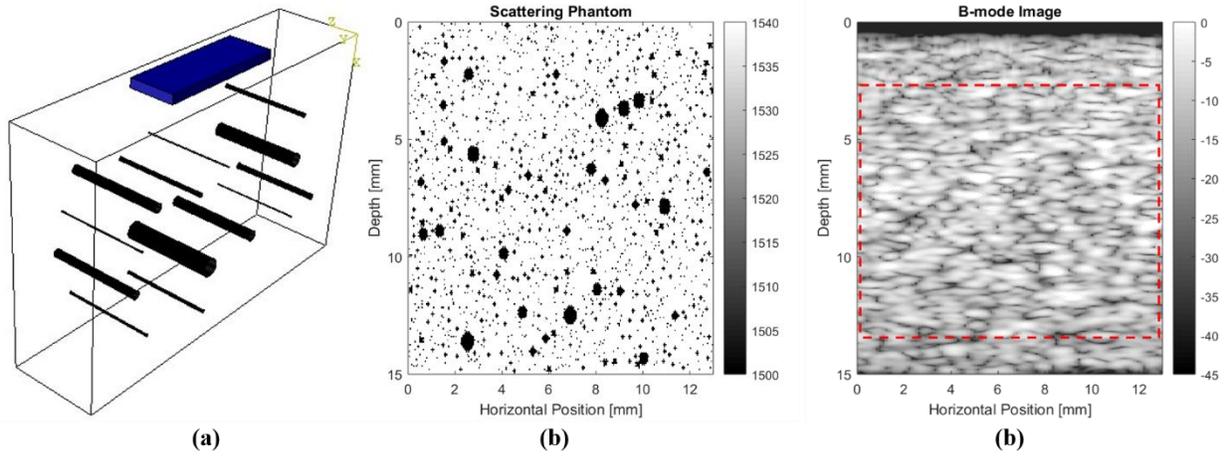

**Fig. 3** (a) 3D orientation of the transducer and random cylindrical scatterers in the simulation domain. Only a few cylindrical branches are shown here to clarify their orientation as perpendicular to the axial propagation of the interrogating pulse. (b) One realization of a random distribution of weak cylindrical scatterers of various diameters following a power law (fractal) function. (c) Resulting 4 MHz B-scan demonstrating speckle pattern.

**Fig. 4** contains the histograms from left to right of the amplitudes (Burr), intensity (Lomax) and log amplitude (logistic) distributions. In each case, the estimated $b$ parameters are near 3.5, higher than the simulated $b = 2.5$. This is expected since in this simulation the estimated number of cylindrical cross sections per sample volume of the interrogating pulse is near 2.5, so the complex addition of Burr phasors acts to increase the power law above its reference value.

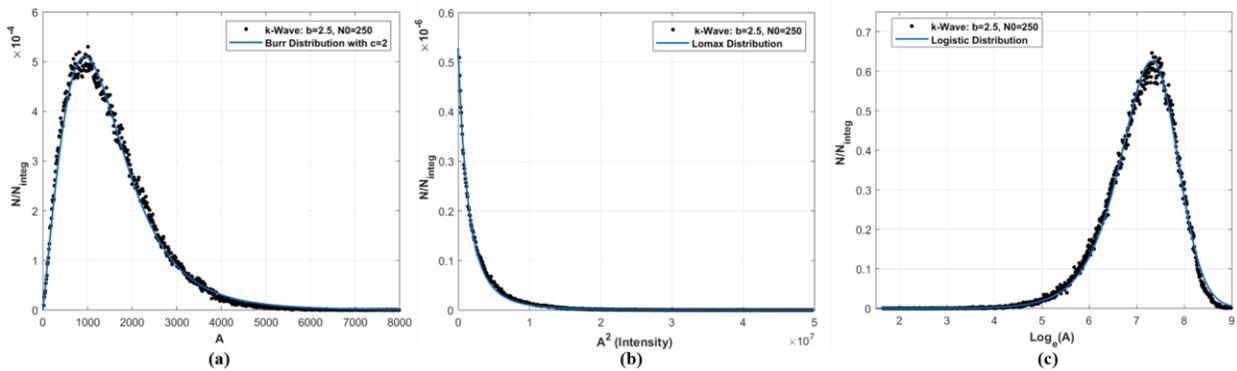

**Fig. 4** Histogram curve fitting from speckle in Figure 3: (a) Burr $b = 3.357$, $\lambda = 2247$. Goodness of fit: SSE = 9.879, $R^2$=0.995. (b) Lomax $b = 3.616$, $\lambda = 5.14e+06$. Goodness of fit: SSE = 0.009627, $R^2$=0.9962. (c) Logistic $b = 3.765$, $\lambda = 2494$, Goodness of fit: SSE = 0.1784, $R^2$=0.9949.

Next, rat livers are examined and ROIs selected within the liver at a depth centered around the transmit focus at 11 mm. **Fig. 5** provides the B-scan of a normal liver, and **Fig. 6** illustrates the Burr, Lomax, and logistic fits to the associated histograms. In these cases, the power law parameters are all estimated to be around 3.8. In comparison, a rat liver from the same study but treated with CCl4 so as to create fibrosis, is shown in **Fig. 7**. The corresponding histograms and theoretical curve fits are shown in **Fig. 8**, and in this case the estimated $b$ parameters are not identical but range from 4.5 (Burr) to 4.9 (logistic) with the Lomax estimate intermediate at 4.7. All these $b$ estimates are higher than those from normal livers and from the simulations, presumably due to the addition of fibrotic patches into the scattering structures of the liver.

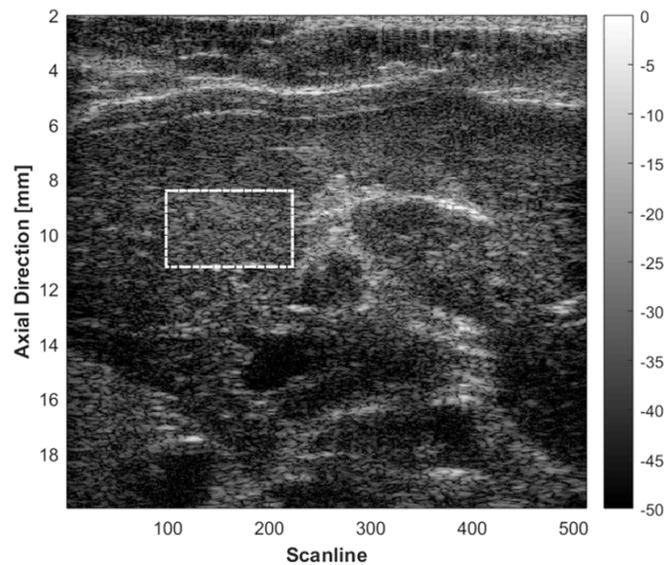

**Fig. 5** Normal liver B-scan.

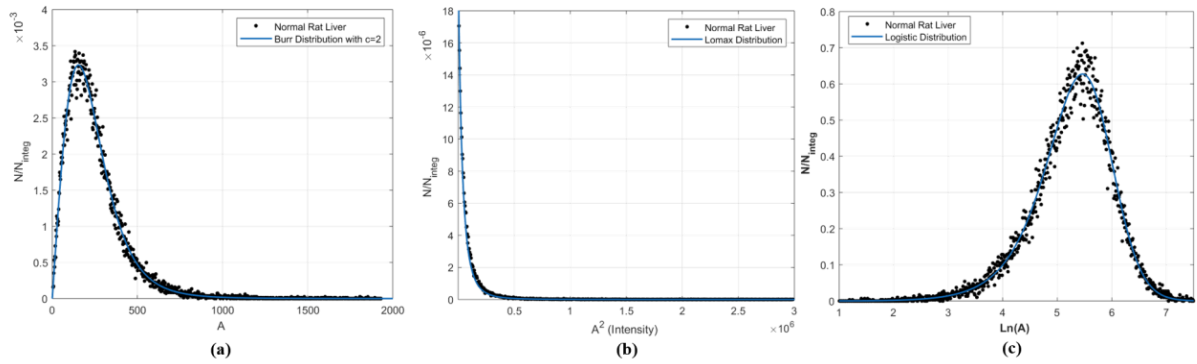

**Fig. 6** Normal liver histograms from speckle in Figure 5. (a) Burr $b=3.828$, $\lambda=396.8$. Goodness of fit: SSE = 5.749, $R^2$=0.9927. (b) Lomax $b=3.77$, $\lambda=1.36\mathrm{e}+05$. Goodness of fit: SSE = 13.22, $R^2$=0.9927. (c) Logistic $b=3.82$, $\lambda=395.7$, Goodness of fit: SSE = 0.4592, $R^2$=0.9874.

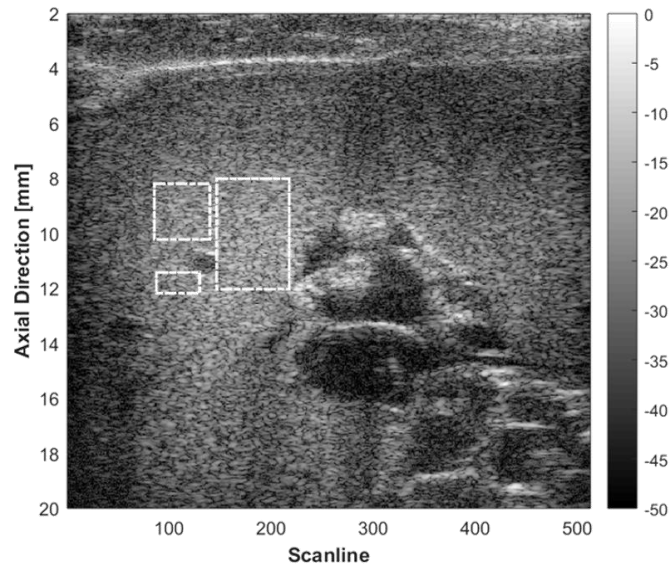

**Fig. 7** Fibrotic liver B-scan.

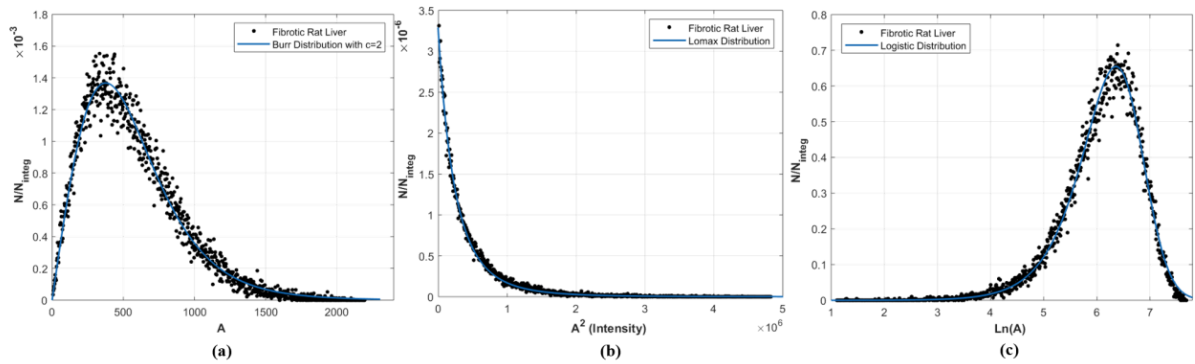

**Fig. 8** Fibrotic liver histograms from speckle in Figure 7. (a) Burr $b=4.467$, $\lambda=1059$. Goodness of fit: SSE = 4.15, $R^2$=0.9801. (b) Lomax $b=4.741$, $\lambda=1.161\mathrm{e}+06$. Goodness of fit: SSE = 1.204, $R^2$=0.9946. (c) Logistic $b=4.955$, $\lambda=1148$, Goodness of fit: SSE = 0.4751, $R^2$=0.9881

## Discussion and Conclusion

There are several key assumptions in the derivation of the PDFs that may limit the applicability of the relationships defined in **Table 1**. First, the convolution model is only an approximation of the complicated wave propagation, but reasonably so for higher f-number focused beams[33]. Furthermore, the fractal model implicitly assumes normal, soft, isotropic tissue with a simple mapping from cylinders of radius $a$ to echo amplitude $A$; the relationship is linear in intensity and single-valued over a certain range of radius compared to wavelength. This is a gross approximation, the precise details depend on the exact nature of the bandpass ultrasound pulse, but the simplified function allows straightforward transformation of probabilities. This raises the possibility that more general forms of the Burr distribution (significantly the three-parameter form of the PDF) may be useful and should be investigated further. Also, fractal models are self-similar across a wide range of scales, however any organ will have limits on the largest and smallest vessels. These limits may influence the statistics of speckle depending on the wavelength of the ultrasound pulse employed, and will require additional consideration. Another limitation is that the model implicitly assumes independent cylindrical scatterers, whereas in reality the branching vasculature is arranged in an orderly manner where each generation originates in a previous generation of vessels. The effect of this on statistics requires further analysis.

The issue of the relative merits of the three main PDFs (Burr, Lomax, and logistic) is a rich area for discussion. Since these have extensive use in the statistics literature, their behaviors are well known in terms of moments, characteristic functions, and estimators of parameters, and a lengthy catalogue of these is beyond the scope of the current discussion. However, the long tail inherent in these distributions places them all in a speckle signal-to-noise ratio of less than the Rayleigh 1.91 theoretical mean to standard deviation.[9] In our investigations, using a minimum mean squared error two-parameter curve fit of different data to each of these, we note that the Lomax distribution for intensity was sometimes the outlier with an elevated $b$ estimate compared to others or compared to the baseline value used in simulations. This may be due to the pronounced concentration of the Lomax histogram near the smallest values of intensity (see **Fig. 4**, **6**, and **8**, middle panel), creating relative insensitivity in the curve fit to the tail of the distribution.

There is an interesting historical twist to these PDFs in that they were originally explored without any reference to ultrasound pulse echo physics. Instead, most of these are associated with economics, income distribution, and complex system lifetimes. The tie between these fields originates with the power law distribution, one of the most ubiquitous laws in natural and human phenomenon.[60] For example in the study of income distribution, a typical country would find many poor people and few rich people. In our ultrasound model, we have many small vessel branches and few large branches. Power law mathematics runs through the core formulas in both fields and then propagates through derived PDFs. Thus, we benefit from the significant work done since the 1940s in fields unrelated to ultrasound. The application of these PDFs to scans from a variety of normal and diseased tissues remains for further investigations.